\documentclass[10pt,conference,letterpaper]{IEEEtran}
\usepackage{times,amsmath}
\usepackage{amsfonts}
\usepackage{amssymb}
\usepackage{amsthm}
\usepackage{algorithm}
\usepackage{algpseudocode}
\usepackage{cite}
\usepackage{graphicx}
\usepackage{epstopdf}
\usepackage[caption=false,font=footnotesize]{subfig}
\usepackage{fixltx2e}

\title{AMMSE Optimization for Multiuser MISO Systems with Imperfect CSIT and Perfect CSIR}
\author{
\IEEEauthorblockN{Hamdi Joudeh\IEEEauthorrefmark{1} and Bruno Clerckx\IEEEauthorrefmark{1}\IEEEauthorrefmark{2}}
\fontsize{9}{9}\upshape
\IEEEauthorrefmark{1} Department of Electrical and Electronic Engineering, Imperial College London, United Kingdom \\
\IEEEauthorrefmark{2} School of Electrical Engineering, Korea University, Seoul, Korea \\
\fontsize{9}{9}\selectfont\ttfamily\upshape
Email: \{hamdi.joudeh10, b.clerckx\}@imperial.ac.uk
}
\begin{document}
\maketitle



\begin{abstract}
In this paper, we consider the design of robust linear precoders for MU-MISO systems where users have perfect Channel State Information (CSI) while the BS has partial CSI.
In particular, the BS has access to imperfect estimates of the channel vectors, in addition to the covariance matrices of the estimation error vectors.
A closed-form expression for the Average Minimum Mean Square Error (AMMSE) is obtained using the second order Taylor Expansion. This approximation is used to formulate two fairness-based robust design problems: a maximum AMMSE-constrained problem and a power-constrained problem.
We propose an algorithm based on convex optimization techniques to address the first problem, while the second problem is tackled by exploiting the close relationship between the two problems, in addition to their monotonic natures.
\end{abstract}

%


\section{Introduction}
The utilization of multiple antennas at the Base Station (BS) combined with simple single-antenna mobile devices could tremendously increase the spectral efficiencies of wireless networks \cite{Viswanath2003}.
However, higher restrictions are imposed, particularly in the Downlink (DL) mode where highly accurate Channel State Information (CSI) is required at the BS. While the ability to provide accurate CSI at the Transmitter (CSIT) remains questionable, considerable work has been done to increase the robustness of transmission schemes initially designed assuming perfect CSI \cite{Shenouda2008,Bogale2013,Joham2010,Gonzalez-Coma2013,Vucic2009}. The robust design problem formulation is highly influenced by the nature of the CSI uncertainty, that varies depending on the context in which it occurs.
%
Two main models have emerged to quantify this uncertainty: the stochastic-uncertainty model \cite{Shenouda2008,Bogale2013,Joham2010,Gonzalez-Coma2013} and the bounded-uncertainty model \cite{Shenouda2008,Vucic2009}. In this paper, we consider robust linear precoding design for Multiuser (MU) Multiple-Input Single-Output (MISO) systems where CSIT uncertainty is modeled stochastically. Particularly, the channel estimation error vectors are assumed to have Gaussian distributed entries. The performance metric considered is the Average Mean Square Error (AMSE), i.e. the expectation of the MSE over the distribution of the channel estimation error.

A similar setup was considered in \cite{Shenouda2008} where the sum AMSE was minimized subject to a total BS power constraint.
While this yields an improved overall performance across users, it does not guarantee fairness. In \cite{Bogale2013}, AMSE fairness-based designs for MU Multiple-Input Multiple-Output (MIMO) systems were proposed.
However, it is assumed in both \cite{Shenouda2008} and \cite{Bogale2013} that CSIT and CSIR have identical uncertainties. While this assumption simplifies the AMSE based transceiver design, it ignores the fact that CSIR is likely to have higher quality.
An AMSE UL-DL duality assuming imperfect CSIT and perfect CSIR was established in \cite{Joham2010}. Based on this duality, an algorithm was proposed to minimize the sum AMSE, where Monte-Carlo integration was applied to calculate expectations that depend on perfect CSI.
A similar problem was addressed in \cite{Bashar2014}, where various approximations were used instead of Monte-Carlo integration.

\emph{Contribution}: For MU-MISO systems with imperfect CSIT and perfect CSIR, we derive a closed-form expression for a Taylor approximation of the Average Minimum Mean Square Error (AMMSE), i.e. the AMSE obtained when MMSE receivers are applied. This approximation is used to formulate two fairness-based robust design problems:
\vspace{-0.5mm}
\begin{enumerate}
  \item The maximum AMMSE-constrained power minimization problem referred to as $\mathcal{P}$.
  \item The power-constrained maximum AMMSE minimization problem referred to as $\mathcal{A}$.
\end{enumerate}
\vspace{-0.5mm}
We propose a fast-converging algorithm based on recursive convex optimization, that solves the non-convex problem $\mathcal{P}$. Furthermore, $\mathcal{A}$ is solved by exploiting its relationship with $\mathcal{P}$; an approach inspired by the work in \cite{Wiesel2006}.

Problem $\mathcal{P}$ was addressed in \cite{Gonzalez-Coma2013} where the authors propose an algorithm based on the ideas in \cite{Joham2010}. However, this algorithm inherits the shortcoming of the approach in \cite{Joham2010}, i.e. expectations are calculated in each iteration via Monte-Carlo integration. Furthermore, convergence could be very slow due to random initialization and limited per-iteration improvement. Results comparing the algorithm propose in this paper to the algorithm in \cite{Gonzalez-Coma2013} are given in Section \ref{Section_Numerical Results}. The rest of the paper is organized as follows: the system model is introduced in Section \ref{Section_System Model}. The AMMSE approximation is derived in Section \ref{Section_AMSE Approximation}. Algorithms that solve $\mathcal{P}$ and $\mathcal{A}$ are proposed in Section \ref{Section_AMSE Robust Beamforming}. Simulation results are presented in Section \ref{Section_Numerical Results} and Section \ref{Section_Conclusion} concludes the paper

\emph{Notation}: Boldface uppercase letters denote matrices, boldface lowercase letters denote column vectors and standard letters denote scalars. The superscrips $(\cdot)^{T}$ and $(\cdot)^{H}$ denote transpose and conjugate-transpose (Hermitian) operators, respectively. $\mathrm{tr}(\cdot)$, $\mathrm{rank}(\cdot)$ and $\|\cdot\|$ are the trace, rank and Euclidian norm operators, respectively. $\mathrm{E}_{x}\{\cdot\}$ denotes the expectation w.r.t the random variable $x$. Finally, $\mathbf{X}\succeq0$ denotes that $\mathbf{X}$ is a Hermitian positive semidefinite matrix.
\section{System Model}
\label{Section_System Model}
We consider a BS equipped with $N_t\geq K$ antennas serving $K$ active single-antenna users. The vector of zero-mean mutually uncorrelated complex data symbols intended for the $K$ receivers is given as $\mathbf{s}=[s_{1},\ldots,s_{K}]^{T} \in\mathbb{C}^{K}$ where $\mathrm{E}\{\mathbf{s}\mathbf{s}^{H}\}=\mathbf{I}$. $\mathbf{s}$ is linearly precoded into the transmit vector $\mathbf{x}\in\mathbb{C}^{N_{t}}$ given as
\vspace{-3mm}
\begin{equation}\label{Eq_x}
  \mathbf{x}=\mathbf{Ps}=\sum_{i=1}^{K}\mathbf{p}_{i}s_{i}
\end{equation}
\vspace{-0.5mm}
where $\mathbf{P}=[\mathbf{p}_{1},\ldots,\mathbf{p}_{K}]$ is the precoding matrix and $\mathbf{p}_{i}\in\mathbb{C}^{N_{t}}$ is the precoding vector consisting of the beamforming weights for the $i$th user. The total transmit power is denoted as $P_{t}$ from which the transmit power constraint could be written as $\mathrm{E}\{\mathbf{x}^{H}\mathbf{x}\} = \mathrm{tr}(\mathbf{P}^{H}\mathbf{P})\leq P_{t}$.
For the $k$th user, the received signal denoted by $y_{k}$ can be written as
\vspace{-1mm}
\begin{equation}\label{Eq_yk}
    y_{k}=\mathbf{h}_{k}^{H}\mathbf{x}+n_{k}
\end{equation}
\vspace{-0.5mm}
where $\mathbf{h}_{k} \in \mathbb{C}^{N_{t}}$ is the narrow-band channel impulse response vector between the $k$th user and the BS. $n_{k} \thicksim \mathcal{CN} ( 0 , \sigma^{2}_{n_{k}} )$ is the Additive White Gaussian Noise (AWGN) at the $k$th user receiver with variance $\sigma_{n_{k}}^{2}$. Throughout the paper, it is assumed that the noise variance is equal across all users i.e. $\sigma_{n_{k}}^{2}=\sigma_{n}^{2}, \ \forall k$.
To obtain an estimate of the intended symbol, each user applies a scalar equalizer $g_{k}$ to its received signal such that $\hat{s}_k=g_{k}y_{k}$.
The Mean Square Error (MSE) at the output of the $k$th receiver is given as
\vspace{-1mm}
\begin{align}
  \varepsilon_{k} =& \ \mathrm{E}_{\mathbf{s},n_{k}}\{|\hat{s}_k - s_k|^{2}\} \\
  \label{Eq_MSE_k}
                 =& \ |g_{k}|^{2}T_{k}-\mathbf{p}_{k}^{H}\mathbf{h}_{k}g_{k}^{H}-g_{k}\mathbf{h}_{k}^{H}\mathbf{p}_{k}+1
\end{align}
where $ T_{k} = \sum_{i=1}^{K} \mathbf{p}_{i}^{H}\mathbf{h}_{k} \mathbf{h}^{H}_{k} \mathbf{p}_{i} + \sigma_{n}^{2}$.
The Minimum Mean Square Error (MMSE) receiver can be calculated by setting the first derivative of $\varepsilon_{k}$, with respect to $g_{k}$, to zero. This yields: $g_{k}(\mathbf{h}_{k}) = \mathbf{p}_{k}^{H}\mathbf{h}_{k}T^{-1}_{k}$.
The notation $g_{k}(\mathbf{h}_{k})$ is used to emphasise the dependency on the channel state $\mathbf{h}_{k}$. In a block-fading channel (which stays constant over a frame of symbols), the $k$th user can calculate $T_{k}$ empirically as the average received signal plus noise power i.e. $T_{k}=\mathrm{E}_{\mathbf{s},n_{k}}\{|y_{k}|^{2}\}$. Furthermore, the scalar effective channel $\mathbf{p}_{k}^{H}\mathbf{h}_{k}$ could be estimated via DL training, from which $g_{k}(\mathbf{h}_{k})$ is calculated.  Plugging $g_{k}(\mathbf{h}_{k})$ into (\ref{Eq_MSE_k}), the MMSE can be written as
\vspace{-1mm}
\begin{align}
  \varepsilon^{\mathrm{MMSE}}_{k} =& \ 1-T^{-1}_{k}R_{k}
  \label{Eq_M_MSE_k}
\end{align}
\vspace{-0.5mm}
where $R_{k}=\mathbf{p}_{k}^{H}\mathbf{h}_{k}\mathbf{h}_{k}^{H}\mathbf{p}_{k}$.
\subsection{CSIT Uncertainty}
The channel vector of the $k$th user can be written as
\vspace{-1mm}
\begin{equation}\label{Eq_hk}
  \mathbf{h}_k= \hat{\mathbf{h}}_{k} + \tilde{\mathbf{h}}_{k}
\end{equation}
\vspace{-0.5mm}
where $\hat{\mathbf{h}}_{k}$ and $\tilde{\mathbf{h}}_{k}$ denote the transmitter-side channel estimate and the channel estimation error, respectively. In this work, the channel estimation error is described statistically, i.e. the entries of $\tilde{\mathbf{h}}_{k}$ are i.i.d Zero-Mean Circularly Symmetric Complex Gaussian (ZMCSCG) with $\mathrm{E}\{\tilde{\mathbf{h}}_{k}\tilde{\mathbf{h}}_{k}^{H}\}=\sigma^{2}_{e_{k}}\mathbf{I}$. $\{\sigma^{2}_{e_{k}}\}_{k=1}^{K}$ and $\{\hat{\mathbf{h}}_{k}\}_{k=1}^{K}$ are assumed to be known by the transmitter, i.e. the BS has first and second order statistics of the channel where
$\mathbf{h}_{k} \thicksim \mathcal{CN} \big( \hat{\mathbf{h}}_{k} , \mathbf{R}_{e_{k}}  \big)$ and $\mathbf{R}_{e_{k}}=\sigma^{2}_{e_{k}}\mathbf{I} $.
Furthermore, it is important to note that all derivations in this paper can be extended to arbitrary $\{\mathbf{R}_{e_{k}}\}_{k=1}^{K}$.
\section{AMMSE and its Approximation}
\label{Section_AMSE Approximation}
Given the statistical nature of the partial CSIT, the optimization of the precoding vectors $\{\mathbf{p}_{k}\}_{k=1}^{K}$ at the BS is carried out in terms of the AMMSE. The AMMSE for the $k$th user is denoted by $\bar{\varepsilon}^{\mathrm{MMSE}}_{k}$ and can be written as
\vspace{-0.5mm}
\begin{equation}\label{Eq_AMSE_k}
  \bar{\varepsilon}^{\mathrm{MMSE}}_{k} = 1 - \mathrm{E}_{{\mathbf{h}}_{k} | \hat{{\mathbf{h}}}_{k}}\{  T_{k}^{-1}R_{k} \}.
\end{equation}
\vspace{-0.5mm}
For notational brevity, $\bar{\varepsilon}^{\mathrm{MMSE}}_{k}$ will be referred to as $\bar{\varepsilon}_{k}$ in the rest of the paper where the use of MMSE receivers is implicit.
Furthermore, $\mathrm{E}_{{\mathbf{h}}_{k} | \hat{{\mathbf{h}}}_{k} }\{\cdot\}$ will be referred to as $\mathrm{E}\{\cdot\}$. The MMSE and the Signal to Interference plus Noise Ratio (SINR) are related such that $\gamma_{k}=\frac{1-\varepsilon_{k}}{\varepsilon_{k}}$, where $\gamma_{k}$ denotes the $k$th user's SINR.
This implies that guaranteeing an AMMSE ensures a minimum average rate \cite{Gonzalez-Coma2013}, and minimizing AMMSE is equivalent to maximizing a lowerbound of the average rate, i.e. $\mathrm{E}\{ \log_{2}(1+\gamma_{k}) \} = \mathrm{E}\{ - \log_{2}(\varepsilon_{k}) \} \geq - \log_{2}( \bar{\varepsilon}_{k})$.
Unfortunately, finding an exact closed-form expressions for (\ref{Eq_AMSE_k}) is not easy.
This difficulty can be addressed by following the assumption in \cite{Shenouda2008}, i.e. ignoring the better quality of CSIR and assuming that it is identical to CSIT. A closed-form expression for AMMSE could be obtained and applied to formulate $\mathcal{P}$ and $\mathcal{A}$. However, ignoring CSIR yields a degraded performance as we demonstrate in the next subsection. To account for perfect CSIR, we propose a close-form expression for an approximation of (\ref{Eq_AMSE_k}).
\subsection{The Ignorant Approach}\label{Sub_Section_Ignorant_AMSE}
Using the available CSIT, both precoders and receivers are optimized at the BS, which informs each user of its corresponding receiver \cite{Shenouda2008}.
In this case, the $k$th receiver is given as $\hat{g}_{k}(\hat{\mathbf{h}}_{k},\sigma^{2}_{e_{k}})= \mathbf{p}_{k}^{H}\hat{\mathbf{h}}_{k}\bar{T}^{-1}_{k}$
where
\vspace{-1mm}
\begin{equation}
\label{Eq_Tk_bar}
  \bar{T}_{k} = \mathrm{E}\{ {T}_{k} \} = \sum_{i=1}^{K} \mathbf{p}_{i}^{H}(\hat{\mathbf{h}}_{k} \hat{\mathbf{h}}^{H}_{k}+\sigma^{2}_{e_{k}}\mathbf{I}) \mathbf{p}_{i} +   \sigma_{n}^{2}.
\end{equation}
\vspace{-0.5mm}
$\hat{g}_{k}(\hat{\mathbf{h}}_{k},\sigma^{2}_{e_{k}})$, which is clearly a function of imperfect CSI, is obtained by minimizing the expectation of (\ref{Eq_MSE_k}), i.e. $\mathrm{E}\{\varepsilon_{k}\}$. The resulting ignorant AMMSE can be written in closed-form as
\vspace{-0.5mm}
\begin{equation}\label{Eq_AMSE_k_Ignorant}
\hat{\varepsilon}_{k} = 1-\bar{T}_{k}^{-1}(\mathbf{p}_{k}^{H}\hat{\mathbf{h}}_{k}\hat{\mathbf{h}}_{k}^{H}\mathbf{p}_{k}).
\end{equation}
\newcounter{Proposition_Counter} 
\newcounter{Remark_Counter} 
\newtheorem{Remark_AMSE_UB}[Remark_Counter]{Remark}
\begin{Remark_AMSE_UB}\label{Remark_AMSE_UB}
\vspace{-0.5mm}
\textnormal{
For any given $\{\mathbf{p}_{k}\}_{k=1}^{K}$, we could write
\vspace{-1mm}
\begin{align}
  \nonumber
  \bar{\varepsilon}_{k}=& \ 1- \mathrm{E} \big\{T_{k}^{-1} \mathbf{p}_{k}^{H} (\hat{\mathbf{h}}_{k}\hat{\mathbf{h}}_{k}^{H} + \tilde{\mathbf{h}}_{k}\tilde{\mathbf{h}}_{k}^{H} + \hat{\mathbf{h}}_{k}\tilde{\mathbf{h}}_{k}^{H} + \tilde{\mathbf{h}}_{k}\hat{\mathbf{h}}_{k}^{H} )\mathbf{p}_{k} \big\}   \\
  \label{Eq_AMSE_UB_1}
                        \leq& \ 1- \mathrm{E} \big\{T_{k}^{-1} \big\} \mathbf{p}_{k}^{H} \hat{\mathbf{h}}_{k} \hat{\mathbf{h}}_{k}^{H} \mathbf{p}_{k}\\
  \label{Eq_AMSE_UB_2}
                        \leq& \ 1- \bar{T}_{k}^{-1} \mathbf{p}_{k}^{H} \hat{\mathbf{h}}_{k} \hat{\mathbf{h}}_{k}^{H} \mathbf{p}_{k} \\
   \nonumber
                        = & \ \hat{\varepsilon}_{k}
\end{align}
where (\ref{Eq_AMSE_UB_1}) follows from the non-negativity of the terms $T_{k}^{-1}$, $\mathbf{p}_{k}^{H}\tilde{\mathbf{h}}_{k}\tilde{\mathbf{h}}_{k}^{H}\mathbf{p}_{k}$ and $\mathbf{p}_{k}^{H}(\hat{\mathbf{h}}_{k}\tilde{\mathbf{h}}_{k}^{H}+\tilde{\mathbf{h}}_{k}\hat{\mathbf{h}}_{k}^{H})\mathbf{p}_{k}$, 
and (\ref{Eq_AMSE_UB_2}) follows from Jensen's inequality.  Equality in (\ref{Eq_AMSE_UB_2}) holds for $\sigma^{2}_{e_{k}}=0$.
}
\end{Remark_AMSE_UB}
\vspace{-2mm}
\newtheorem{Remark_AMMSE_Scaling}[Remark_Counter]{Remark}
\begin{Remark_AMMSE_Scaling}\label{Remark_AMMSE_Scaling}
\textnormal{For a set of precoding vectors $\{c\mathbf{p}_{k}\}_{k=1}^{K}$ where $c$ is a positive power-scaling factor, $\{\bar{\varepsilon}_{k}\}_{k}^{K}$ and $\{\hat{\varepsilon}_{k}\}_{k}^{K}$ are monotonically non-increasing in $c$. This is evident from plugging $\{c\mathbf{p}_{k}\}_{k=1}^{K}$ into (\ref{Eq_AMSE_k}) and (\ref{Eq_AMSE_k_Ignorant}).}
\end{Remark_AMMSE_Scaling}
\vspace{-3mm}
\newcounter{Corollary_Counter} 
\newtheorem{Corollary_Ignorant_Aware}[Corollary_Counter]{Corollary}
\begin{Corollary_Ignorant_Aware}\label{Corollary_Ignorant_Aware}
\textnormal{The ignorant approach yields a degraded performance (higher AMMSEs or power) compared to the an aware approach, that takes into account the perfect CSIR.
For a power-constrained problem (e.g. $\mathcal{A}$), (\ref{Eq_AMSE_UB_2}) holds even if $\{\mathbf{p}_{k}\}_{k=1}^{K}$ were ignorant precoders, i.e. optimally designed w.r.t $\{\hat{\varepsilon}_{k}\}_{k=1}^{K}$. For an AMMSE constrained problem (e.g. $\mathcal{P}$), Remark \ref{Remark_AMMSE_Scaling} implies that aware-optimization could achieve the same AMMSEs for less power compared to ignorant-optimization by using a down-scaled version of the optimum ignorant precoders.
Moreover, even if we assume that perfect CSIR is utilized by users
and ignorant AMMSEs are only used as optimization metric at the BS, this corresponds to using upperbounds of the
AMMSEs which can be very loose under certain channel conditions \cite{Bashar2014}.
}
\end{Corollary_Ignorant_Aware}
\vspace{-2mm}
\subsection{AMMSE Taylor Approximation}
The expectation of a ratio of two random variables could be approximated using the Taylor expansion \cite{Rice2009}:
\vspace{-2mm}
\newcounter{Lemma_Counter} 
\newtheorem{Lemma_Taylor_Expansion}[Lemma_Counter]{Lemma}
\begin{Lemma_Taylor_Expansion}\label{Lemma_Taylor_Expansion}
For two random variables $x$ and $y$ with expectations $\mathrm{E}\{x\}=\bar{x}$ and $\mathrm{E}\{y\}=\bar{y}$, and $y\neq0$, we could write
\vspace{-1mm}
\begin{equation}\label{Eq_Taylor}
\mathrm{E}\bigg\{\frac{x}{y}\bigg\} \approx \frac{\bar{x}}{\bar{y}} + \sum_{i = 1}^{2(N-1)} (-1)^{i} \frac{\bar{x}\mu_{0,i}+\mu_{1,i}}{\bar{y}^{i+1}}
\end{equation}
\vspace{-0.5mm}
where $N$ is the order of the Taylor expansion and
\vspace{-1mm}
\begin{equation}\label{Eq_Taylor_Moment}
\mu_{m,n} = \mathrm{E}\{ (x-\bar{x})^{m} (y-\bar{y})^{n} \}.
\end{equation}
\end{Lemma_Taylor_Expansion}
\vspace{-2mm}
The accuracy of the approximation in (\ref{Eq_Taylor}) increases as $N$ grows larger where $N\rightarrow\infty$ yields infinite accuracy. An approximation of (\ref{Eq_AMSE_k}) is denoted by $\bar{\varepsilon}_{k}^{(N)}$, where $N$ is the order of the Taylor expansion of $\mathrm{E}\{T_{k}^{-1}R_{k}\}$. To maintain tractability, we consider the second-order Taylor expansion which could be written as \cite{VanKempen2000}
\vspace{-0.5mm}
\begin{align}
  \nonumber
  \bar{\varepsilon}_{k}^{(2)} =&  1-\bigg(\frac{\bar{R}_{k}}{\bar{T}_{k}} - \frac{\text{cov}\{R_{k},T_{k}\}}{\bar{T}_{k}^{2}} + \frac{\bar{R}_{k} \text{var}\{T_{k}\}}{\bar{T}_{k}^{3}}  \bigg) \\
  \label{Eq_AMSE_k_Approx_2}
                        =&  1 \! - \! \frac{\bar{R}_{k}}{\bar{T}_{k}} \! + \! \frac{\mathrm{E} \! \{ \! R_{k} \! T_{k} \! \} \! - \! \bar{R}_{k} \! \bar{T}_{k}}{\bar{T}_{k}^{2}} \! - \! \frac{\bar{R}_{k} \! (\mathrm{E} \! \{ \! T_{k}^{2} \! \} \! - \! \bar{T}_{k}^{2})}{\bar{T}_{k}^{3}}
\end{align}
\vspace{-1mm}
where
\vspace{-1mm}
\begin{equation}
\label{Eq_Rk_bar}
  \bar{R}_{k} = \mathrm{E}\{ {R}_{k} \} = \mathbf{p}_{k}^{H}(\hat{\mathbf{h}}_{k} \hat{\mathbf{h}}^{H}_{k}+\sigma^{2}_{e_{k}}\mathbf{I}) \mathbf{p}_{k}.
\end{equation}
\vspace{-1mm}
Each of the terms $\mathrm{E}\{R_{k}T_{k}\}$ and $\mathrm{E}\{T_{k}^{2}\}$ in (\ref{Eq_AMSE_k_Approx_2}) is an expectation of a product of two quadratic forms in a random vector $\mathbf{h}_{k} \thicksim \mathcal{CN} \big( \hat{\mathbf{h}}_{k} , \sigma^{2}_{e_{k}}\mathbf{I}  \big)$. Closed-form expressions can be obtained using the following lemma:
\vspace{-1mm}
\newtheorem{Lemma_E_Complex_Quadratic_non_zero_mean}[Lemma_Counter]{Lemma}
\begin{Lemma_E_Complex_Quadratic_non_zero_mean}\label{Lemma_E_Complex_Quadratic_non_zero_mean}
For a complex gaussian vector $\mathbf{x}\sim\mathcal{CN}(\hat{\mathbf{x}},\mathbf{C})$ given as $\mathbf{x}=\hat{\mathbf{x}}+\tilde{\mathbf{x}}$ where $\hat{\mathbf{x}}$ is the mean and $\tilde{\mathbf{x}}$ has ZMCSCG entries, and two quadratic forms defined as $Q_{1}=\mathbf{x}^{H}\mathbf{A}\mathbf{x}$ and $Q_{2}=\mathbf{x}^{H}\mathbf{B}\mathbf{x}$ where $\mathbf{A}, \mathbf{B}\succeq 0$,
we have
\vspace{-2mm}
\begin{multline}
\label{Eq_E_Q1Q2}
\mathrm{E}\{Q_{1}Q_{2}\} =  \hat{\mathbf{x}}^{H}\mathbf{ACB}\hat{\mathbf{x}}+\hat{\mathbf{x}}^{H}\mathbf{BCA}\hat{\mathbf{x}}+\textnormal{tr}(\mathbf{ACBC})\\
+
(\textnormal{tr}(\mathbf{AC})+\hat{\mathbf{x}}^{H}\mathbf{A}\hat{\mathbf{x}})(\textnormal{tr}(\mathbf{BC})+\hat{\mathbf{x}}^{H}\mathbf{B}\hat{\mathbf{x}}).
\end{multline}
\end{Lemma_E_Complex_Quadratic_non_zero_mean}
\vspace{-2mm}
A sketch of the proof is provided in the Appendix. Using Lemma \ref{Lemma_E_Complex_Quadratic_non_zero_mean}, we could write
\vspace{-1mm}
\begin{align}
\nonumber
\mathrm{E}\{R_{k}T_{k}\} =& \ \sigma_{e_{k}}^{2}\hat{\mathbf{h}}_{k}^{H}\mathbf{Q}_{k}\mathbf{Q}\hat{\mathbf{h}}_{k}+
\sigma_{e_{k}}^{2}\hat{\mathbf{h}}_{k}^{H}\mathbf{Q}\mathbf{Q}_{k}\hat{\mathbf{h}}_{k}\\
\label{Eq_E_Rk_Tk}
 & + (\sigma_{e_{k}}^{2})^{2}\mathrm{tr}(\mathbf{Q}_{k}\mathbf{Q})+\bar{R}_{k}\bar{T}_{k} \\
 \label{Eq_E_Tk2}
\mathrm{E}\{T_{k}^{2}\} =& \ 2\sigma_{e_{k}}^{2}\hat{\mathbf{h}}_{k}^{H}\mathbf{Q}^{2}\hat{\mathbf{h}}_{k}+
  (\sigma_{e_{k}}^{2})^{2}\mathrm{tr}(\mathbf{Q}^{2})+\bar{T}_{k}^{2}
\end{align}
where $\mathbf{Q}_{k}=\mathbf{p}_{k}\mathbf{p}_{k}^{H}$ and $\mathbf{Q}=\sum_{i=1}^{K}\mathbf{p}_{i}\mathbf{p}_{i}^{H}$. Before plugging (\ref{Eq_E_Rk_Tk}) and (\ref{Eq_E_Tk2}) back into (\ref{Eq_AMSE_k_Approx_2}), we define $a_{k}$ and $b_{k}$ as follows
\vspace{-1mm}
\begin{align}
\label{Eq_ak}
a_{k} =& \ \hat{\mathbf{h}}_{k}^{H}(\mathbf{Q}_{k}\mathbf{Q}+\mathbf{Q}\mathbf{Q}_{k})\hat{\mathbf{h}}_{k}
  + \sigma_{e_{k}}^{2}\mathrm{tr}(\mathbf{Q}_{k}\mathbf{Q}), \ \forall k \\
\label{Eq_bk}
b_{k} =& \ 2\hat{\mathbf{h}}_{k}^{H}\mathbf{Q}^{2}\hat{\mathbf{h}}_{k}+
  \sigma_{e_{k}}^{2}\mathrm{tr}(\mathbf{Q}^{2}) , \ \forall k
\end{align}
\vspace{-1mm}
from which (\ref{Eq_AMSE_k_Approx_2}) could be written as
\vspace{-1mm}
\begin{align}
 \label{Eq_AMSE_k_Approx_3}
  \bar{\varepsilon}_{k}^{(2)} =&\  1-\frac{\bar{R}_{k}}{\bar{T}_{k}}
  +a_{k}\frac{\sigma_{e_{k}}^{2}}{\bar{T}_{k}^{2}} -
  b_{k}\frac{\sigma_{e_{k}}^{2}\bar{R}_{k}}{\bar{T}_{k}^{3}}   \\
  \label{Eq_AMSE_k_Approx_4}
   =&\ 1-\alpha_{k}\frac{\bar{R}_{k}}{\bar{T}_{k}}
\end{align}
\vspace{-3mm}
where $\alpha_{k}$ is given as
\vspace{-1mm}
\begin{equation}
 \label{Eq_Alpha_k}
  \alpha_{k} = \ 1 - \frac{\sigma_{e_{k}}^{2}}{\bar{T}_{k}^{2} \bar{R}_{k}}(a_{k}\bar{T}_{k} - b_{k}\bar{R}_{k}).
\end{equation}
It is clear that (\ref{Eq_AMSE_k_Approx_4}) reduces to $\bar{\varepsilon}_{k}^{(1)}$ if $\alpha_{k}$ was replaced by $1$, while $\hat{\varepsilon}_{k}$ will be obtained if it was replaced by $\big(|\hat{\mathbf{h}}_{k}^{H}\mathbf{p}_{k}|^{2}/\bar{R}_{k}\big)$.
\section{AMMSE Based Robust Beamforming}
\label{Section_AMSE Robust Beamforming}
In this section, the AMMSE approximation in (\ref{Eq_AMSE_k_Approx_4}) is used to formulate the robust design problems $\mathcal{P}$ and $\mathcal{A}$. The approach followed to solve those problems is based on the one proposed in \cite{Wiesel2006}. Particularly, the solution of the former problem is based on conic optimization, where the later problem is solved by exploiting the relationship between the two problems, and the monotonic nature of their objective functions.
\subsection{Power Minimization}
 \label{Subsection_Power_Min}
The power minimization problem is denoted by $\mathcal{P}(\bar{\varepsilon}_{0})$ where $0<\bar{\varepsilon}_{0}<1$ is the given worst AMMSE constraint. $\mathcal{P}(\bar{\varepsilon}_{0})$ could be formulated as
\vspace{-1mm}
\begin{align}
\nonumber
 \mathcal{P}(\bar{\varepsilon}_{0}): \  &\underset{\mathbf{p}_{1},\ldots,\mathbf{p}_{K}}{\min} \ \sum_{i=1}^{K}\|\mathbf{p}_{i}\|^{2}  \\
 \label{Eq_Opt_P}
    & \ \ \  \text{s.t.} \ \ \ 1-\alpha_{k}\frac{\bar{R}_{k}}{\bar{T}_{k}} \leq \bar{\varepsilon}_{0},\ \ \forall k\in\{1,\ldots,K\}.
\end{align}
Introducing the non-negative real-valued slack variable $P_{0}$, and transforming the objective and constraints into the Linear Matrix Inequality (LMI) form, (\ref{Eq_Opt_P}) could be rewritten as
\vspace{-1mm}
\begin{align}
\nonumber
 \mathcal{P}(\bar{\varepsilon}_{0}): \  &\underset{P_{0},\mathbf{Q}_{1},\ldots,\mathbf{Q}_{K}}{\min} \ P_{0}   \\
 \nonumber
 & \ \ \ \  \text{s.t.} \ \ \  \sum_{i=1}^{K}\mathrm{tr}(\mathbf{Q}_{i}) \leqslant P_{0}  ,
 \\
\nonumber
 \sum_{i=1}^{K}\!
\mathrm{tr} \!
&
\Big(\! \mathbf{Q}_{i} \! \big(  \hat{\mathbf{h}}_{k} \! \hat{\mathbf{h}}_{k}^{H} \! + \! \sigma_{e_{k}}^{2} \! \mathbf{I} \! \big) \! \Big)\!
+ \! \sigma_{n}^{2} \! \leq \!
\frac{\alpha_{k}}{1\! - \! \bar{\varepsilon}_{0}}
\mathrm{tr} \!  \Big( \! \mathbf{Q}_{k} \! \big(  \hat{\mathbf{h}}_{k} \! \hat{\mathbf{h}}_{k}^{H} \! + \! \sigma_{e_{k}}^{2} \! \mathbf{I} \!  \big) \!  \Big) ,
\\
\nonumber
    & \ \ \ \   \ \ \ \    \  \mathbf{Q}_{k}\succeq 0  ,
\\
    \label{Eq_Opt_P_2}
    & \ \ \ \    \ \ \ \    \  \mathrm{rank}(\mathbf{Q}_{k})= 1  , \ \ \forall k\in\{1,\ldots,K\}.
\end{align}
\vspace{-0.5mm}
Assuming that $\mathcal{P}(\bar{\varepsilon}_{0})$ is feasible, finding the optimal $P_{0}$ in (\ref{Eq_Opt_P_2}) could be very challenging. This is mainly because the second set of constraints are non-linear due to the presence of $\alpha_{k}$, and the rank constraints are non-convex.
To make the problem less complicated, the non-linearity in the second set of constraints could be eliminated by replacing $\{\alpha_{k}\}_{k=1}^{K}$ with fixed, real and non-negative values denoted by $\{\bar{\alpha}_{k}\}_{k=1}^{K}$; e.g. $\bar{\alpha}_{k}$ could be set to $1$ where the AMMSE constraints will be defined in terms of the first-order approximation of $\{\bar{\varepsilon}_{k}\}_{k=1}^{K}$, i.e. $\{\bar{\varepsilon}_{k}^{(1)}\}_{k=1}^{K}$. The linearized problem is denoted by $\mathcal{P}_{l}(\bar{\varepsilon}_{0},\{\bar{\alpha}_{k}\}_{k=1}^{K})$ and takes $1+K$ input arguments. $\mathcal{P}_{l}$ could be made tractable by relaxing the rank constraints. The new linearized and relaxed problem, referred to as $\mathcal{P}_{lr}(\bar{\varepsilon}_{0},\{\bar{\alpha}_{k}\}_{k=1}^{K})$, is convex as it is composed of a linear objective function and a combination of linear and semidefinite constraints. In particular, $\mathcal{P}_{lr}$ is a Semidefinite Program (SDP) which could be solved efficiently using Interior-Point methods \cite{Boyd2004}.

Due to the rank relaxation, the $K$ matrices $\{\mathbf{Q}_{k}\}_{k=1}^{K}$ obtained by solving $\mathcal{P}_{lr}$ will not be rank-1 in general. If they are all rank-1, then the optimum solutions for $\mathcal{P}_{lr}$ and $\mathcal{P}_{l}$ coincide and $\{\mathbf{p}_{k}\}_{k=1}^{K}$ could be obtained directly through eigen decomposition. Otherwise, the power obtained by solving $\mathcal{P}_{lr}$ is a lowerbound for the the optimum objective value of $\mathcal{P}_{l}$. This is due to the fact that relaxation extends the domain of feasible $\{\mathbf{Q}_{k}\}_{k=1}^{K}$ bearing the possibility of a solution with a lower objective compared to the non-relaxed problem.

For each $k$, if $\mathrm{rank}(\mathbf{Q}_{k})>1$, then $\mathbf{p}_{k}$ could be chosen as the principal eigenvector of $\mathbf{Q}_{k}$ or generated using randomization \cite{Luo2010}. However, it is likely that the resulting beamforming vectors will fail to satisfy the AMMSE constraints. In this case, further optimization is required for power reallocation. In the simulations carried out for this paper, it has been observed that solving $\mathcal{P}_{lr}$ always gives rank-1 solutions. Further investigations regarding this observation is left for future work.
\vspace{-4mm}
\begin{algorithm}[H]
\caption{Power Minimization}
\label{Algthm_Power_Opt}
\begin{algorithmic}[1]
\State \textbf{Initialize}: $n\gets 0$, $P_{0}^{(n)}\gets 0 $, $\bar{\alpha}_k^{(n)}\gets1 \ \forall k $
\Repeat
    \State $n\gets n+1$
    \State $P_{0}^{(n)} \gets \mathcal{P}_{lr}(\bar{\varepsilon}_{0},\{\bar{\alpha}_{k}^{(n-1)}\}_{k=1}^{K})$
    \State $\{\mathbf{Q}_{k}^{(n)}\}_{k=1}^{K} \gets \arg \mathcal{P}_{lr}(\bar{\varepsilon}_{0},\{\bar{\alpha}_{k}^{(n-1)}\}_{k=1}^{K})$
    \State $\mathbf{Q}^{(n)} \gets \sum_{i=1}^{i=K}\mathbf{Q}_{i}^{(n)}$
    \State \text{update} $\{\bar{\alpha}_{k}^{(n)}\}_{k=1}^{K}$ \text{using (\ref{Eq_Alpha_k})}
\Until{$|P_{0}^{(n)}-P_{0}^{(n-1)}|<\epsilon_{P}$ \text{or} $n=n_{\max}$ }
\end{algorithmic}
\end{algorithm}
\vspace{-4mm}

Going back to the original power optimization problem $\mathcal{P}$, a solution is proposed which involves solving $\mathcal{P}_{lr}$ recursively over multiple iterations where the values of $\{\bar{\alpha}_{k}\}_{k=1}^{K}$ are updated in each iteration.
Particularly, in the $n$th iteration, $\mathcal{P}_{lr}(\bar{\varepsilon}_{0},\{\bar{\alpha}_{k}^{(n-1)}\}_{k=1}^{K})$ is optimally solved where $\{\bar{\alpha}_{k}^{(n-1)}\}_{k=1}^{K}$ are obtained using the solution of $\mathcal{P}_{lr}$ in the $(n-1)$th iteration. This is carried out until a desired accuracy (specified by $\epsilon_{P}$) is achieved or a maximum number of iterations $n_{\max}$ is reached. The pseudo-code for this method is shown in Algorithm \ref{Algthm_Power_Opt}. Although Algorithm \ref{Algthm_Power_Opt} is not guaranteed to reach a global optimum, simulations show that quick convergence with good performance is achieved.

It is important to highlight that the presence of CSIT uncertainty may impose a feasibility bound on $\bar{\varepsilon}_{0}$.
For perfect CSI i.e. $\sigma_{e_{k}}^{2}=0 \ \forall k$, any $\bar{\varepsilon}_{0} \geq 0$ is feasible. This is directly concluded from ${\gamma}_{0} \leq \infty$ in \cite{Wiesel2006} where ${\gamma}_{0}$ denotes the target SINR. Feasibility for the case where CSIT is imperfect has been addressed in \cite{Gonzalez-Coma2013a}.
Although no closed-form expression has been derived (a QoS region could be obtained through Monte-Carlo integration), it has been observed that $\bar{\varepsilon}_{0}$ is lowerbounded above $0$.
This could be confirmed by plugging the scaled precoding vectors $\{c\mathbf{p}_{k}\}_{k=1}^{K}$ into $\bar{\varepsilon}_{k}^{(2)}$ and driving the transmit power up to infinity, i.e. $c\rightarrow\infty$. Regardless of the structure of $\{\mathbf{p}_{k}\}_{k=1}^{K}$, for non-zero fixed ${\{\sigma_{e_{k}}^{2}}\}_{k=1}^{K}$ that do not scale down with increased power, residual interference terms will bound $\bar{\varepsilon}_{k}^{(2)}$ above $0$.
\subsection{Maximum AMMSE Minimization}
The minimization of the maximum AMMSE problem is referred to as $\mathcal{A}(P_{t})$ where $P_{t}$ denotes the total transmission power constraint. This could be written as
\vspace{-2mm}
\begin{align}
\nonumber
 \mathcal{A}(P_{t}): \ &\underset{\mathbf{p}_{1},\ldots,\mathbf{p}_{K}}{\min} \ \underset{k}{\max} \ \ 1-\alpha_{k}\frac{\bar{R}_{k}}{\bar{T}_{k}}  \\
 \label{Eq_Opt_A}
    & \ \ \ \text{s.t.} \ \ \  \ \  \sum_{i=1}^{K}\|\mathbf{p}_{i}\|^{2} \leq  P_{t}.
\end{align}
\vspace{-0.5mm}
By introducing the non-negative real-valued slack variable $t_{0}$ and adding the constraints $ 1-\alpha_{k}{\bar{T}_{k}}^{-1}{\bar{R}_{k}} \leq t_{0} \ \forall k$, the objective in (\ref{Eq_Opt_A}) could be written as: $\min \ t_{0}$. Following the formulation of (\ref{Eq_Opt_P_2}), (\ref{Eq_Opt_A}) could be rewritten as
\vspace{-1mm}
\begin{align}
\nonumber
 \mathcal{A}(P_{t}): \  &\underset{t_{0},\mathbf{Q}_{1},\ldots,\mathbf{Q}_{K}}{\min} \ t_{0}   \\
 \nonumber
 & \ \ \ \  \text{s.t.} \ \ \  \sum_{i=1}^{K}\mathrm{tr}(\mathbf{Q}_{i}) \leqslant P_{t}  ,
 \\
\nonumber
 \sum_{i=1}^{K}\!
\mathrm{tr} \!
&
\Big(\! \mathbf{Q}_{i} \! \big(  \hat{\mathbf{h}}_{k} \! \hat{\mathbf{h}}_{k}^{H} \! + \! \sigma_{e_{k}}^{2} \! \mathbf{I} \! \big) \! \Big)\!
+ \! \sigma_{n}^{2} \! \leq \!
\frac{\alpha_{k}}{1\! - \! t_{0}}
\mathrm{tr} \!  \Big( \! \mathbf{Q}_{k} \! \big(  \hat{\mathbf{h}}_{k} \! \hat{\mathbf{h}}_{k}^{H} \! + \! \sigma_{e_{k}}^{2} \! \mathbf{I} \!  \big) \!  \Big) ,
\\
\nonumber
    & \ \ \ \   \ \ \ \    \  \mathbf{Q}_{k}\succeq 0  ,
\\
    \label{Eq_Opt_A_2}
    & \ \ \ \    \ \ \ \    \  \mathrm{rank}(\mathbf{Q}_{k})= 1  , \ \ \forall k\in\{1,\ldots,K\}.
\end{align}
\vspace{-0.5mm}
At a first glance, (\ref{Eq_Opt_A_2}) looks similar to (\ref{Eq_Opt_P_2}). However, after careful consideration one could see that $t_{0}$ in the second set of constraints in (\ref{Eq_Opt_A_2}) is an optimization variable,
contrary to $\bar{\varepsilon}_{0}$ in (\ref{Eq_Opt_P_2}) which is a given input parameter.
Same applies to $P_{t}$ and $P_{0}$ where the former is a power constraint input parameter in (\ref{Eq_Opt_A_2}) while the later is the objective variable in (\ref{Eq_Opt_P_2}). $\mathcal{A}_{lr}(P_{t},\{\bar{\alpha}_{k}\}_{k=1}^{K})$ denotes a semi-linearized and relaxed version of $\mathcal{A}$ obtained by  fixing $\{{\alpha}_{k}\}_{k=1}^{K}$ to $\{\bar{\alpha}_{k}\}_{k=1}^{K}$ and relaxing the rank constraints. However, non-linearity is still present in the constraints of $\mathcal{A}_{lr}$ due to the presence of the objective variable $t_{0}$. This could be tackled by exploiting the relationship between $\mathcal{P}_{lr}$ and $\mathcal{A}_{lr}$, and their monotonic nature.
\newtheorem{Proposition_P_A_Inverse_Problems_and_Monotonicity}[Proposition_Counter]{Proposition}
\begin{Proposition_P_A_Inverse_Problems_and_Monotonicity}\label{Proposition_P_A_Inverse_Problems_and_Monotonicity}
\vspace{-1mm}
$\mathcal{P}_{lr}$ and $\mathcal{A}_{lr}$ are inverse problems such that:
\begin{equation}\label{Eq_A_P_Inverse_Problems}
  \mathcal{A}_{lr}\big(\mathcal{P}(\bar{\varepsilon}_{0},\{\bar{\alpha}_{k}\}_{k=1}^{K}), \{\bar{\alpha}_{k}\}_{k=1}^{K}\big)  =  \bar{\varepsilon}_{0}
\end{equation}
\begin{equation}\label{Eq_P_A_Inverse_Problems}
  \mathcal{P}_{lr}\big(\mathcal{A}(P_{t},\{\bar{\alpha}_{k}\}_{k=1}^{K}), \{\bar{\alpha}_{k}\}_{k=1}^{K}\big)  =  P_{t}.
\end{equation}
In addition, the optimum objective of $\mathcal{P}_{lr}$ and $\mathcal{A}_{lr}$ are monotonic such that:
\begin{equation}\label{Eq_P_Monotonicity}
  \mathcal{P}_{lr}(\bar{\varepsilon}_{0},\{\bar{\alpha}_{k}\}_{k=1}^{K}) >  \mathcal{P}_{lr}(\bar{\varepsilon}_{0}^{o},\{\bar{\alpha}_{k}\}_{k=1}^{K}) \Rightarrow \bar{\varepsilon}_{0} < \bar{\varepsilon}_{0}^{o}
\end{equation}
\begin{equation}\label{Eq_A_Monotonicity}
  \mathcal{A}_{lr}(P_{t},\{\bar{\alpha}_{k}\}_{k=1}^{K}) \leq \mathcal{A}_{lr}(P_{t}^{o},\{\bar{\alpha}_{k}\}_{k=1}^{K}) \Rightarrow P_{t} > P_{t}^{o}
\end{equation}
where $\bar{\varepsilon}_{0}$ and $\bar{\varepsilon}_{0}^{o}$ are assumed to be feasible.
\end{Proposition_P_A_Inverse_Problems_and_Monotonicity}
\vspace{-1mm}
\begin{proof}[Proof of Proposition \ref{Proposition_P_A_Inverse_Problems_and_Monotonicity}]
This could be proved by contradiction \cite{Wiesel2006}. Due to the lack of space, the reader is referred to the proof of \cite[Theorem 3]{Wiesel2006}. Contrary to \cite[(67)]{Wiesel2006}, the left-hand side inequality in (\ref{Eq_A_Monotonicity}) is not strict. This is due to the possibility that $1-\bar{\alpha}_{k}{\bar{T}_{k}}^{-1}{\bar{R}_{k}}$ will hit a floor at high SNRs, as shown in Section \ref{Subsection_Power_Min}, causing looseness to the power constraint in $\mathcal{A}_{lr}$.
\end{proof}
\vspace{-1mm}
\newtheorem{Corollary_A_Solution_Bisection}[Corollary_Counter]{Corollary}
\begin{Corollary_A_Solution_Bisection}\label{Corollary_A_Solution_Bisection}
\textnormal{Propositions \ref{Proposition_P_A_Inverse_Problems_and_Monotonicity} implies that $\mathcal{A}_{lr}(P_{0},\{\bar{\alpha}_{k}\}_{k=1}^{K})$ could be optimally solved by carrying out a one-dimensional bisections search over $\bar{\varepsilon}_{0}$ until the minimum $\bar{\varepsilon}_{0}$ that satisfies $\mathcal{P}_{lr}(\bar{\varepsilon}_{0},\{\bar{\alpha}_{k}\}_{k=1}^{K}) \leq P_{0}$ is found.}
\end{Corollary_A_Solution_Bisection}
\vspace{-1mm}
The pseudo-code for the bisections search method is given in Algorithm \ref{Algthm_AMSE_Opt_Bisection} where the value of $\epsilon_{0}$ determines the accuracy of the solution.
\vspace{-2mm}
\begin{algorithm}
\caption{Max. AMMSE Minimization, fixed $\{\bar{\alpha}_{k}\}_{k\!=\!1}^{K}$}
\label{Algthm_AMSE_Opt_Bisection}
\begin{algorithmic}[1]
\State \textbf{Initialize}: $\bar{\varepsilon}_{\min} \gets 0$, $\bar{\varepsilon}_{\max}\gets 1 $
\Repeat
    \State $\bar{\varepsilon}_{0}\gets (\bar{\varepsilon}_{\min}+\bar{\varepsilon}_{\max})/2$
    \State $\bar{P}_{0} \gets \mathcal{P}_{lr}(\bar{\varepsilon}_{0},\{\bar{\alpha}_{k}\}_{k=1}^{K})$ \label{Algthm_AMSE_Opt_Bisection_Step_P}

    \If {$\bar{P}_{0} > {P}_{0}$ or $\mathcal{P}_{lr}(\bar{\varepsilon}_{0},\{\bar{\alpha}_{k}\}_{k=1}^{K})$ is infeasible}
    \State $\bar{\varepsilon}_{\min} \gets \bar{\varepsilon}_{0}$
    \Else
    \State $\bar{\varepsilon}_{\max} \gets \bar{\varepsilon}_{0}$
    \EndIf
\Until{$\bar{\varepsilon}_{\max} - \bar{\varepsilon}_{\min} \leq \epsilon_{0}$}
\State $t_{0} \gets \bar{\varepsilon}_{\max}$
\State $\{\mathbf{Q}_{k}\}_{k=1}^{K} \gets \arg \mathcal{P}_{lr}(\bar{\varepsilon}_{\max},\{\bar{\alpha}_{k}\}_{k=1}^{K})$
\end{algorithmic}
\end{algorithm}
\vspace{-2mm}
The stopping criteria in the bisection search is expressed in terms of $\bar{\varepsilon}_{\min}$ and $\bar{\varepsilon}_{\max}$ rather than $P_{0}$ (as in \cite{Wiesel2006}) due to the possibility of power constraint looseness. Furthermore, $t_{0}$ is set to $\bar{\varepsilon}_{\max}$ and the solution is given by $\arg \mathcal{P}_{lr}(\bar{\varepsilon}_{\max},\{\bar{\alpha}_{k}\}_{k=1}^{K})$, as $\bar{\varepsilon}_{\min}$ could be infeasible.
Following the same recursive approach used to find a solution for $\mathcal{P}$, a solution for problem $\mathcal{A}$ could be obtained using Algorithm \ref{Algthm_AMSE_Opt}. As for Algorithm \ref{Algthm_Power_Opt}, quick convergence and good performance could be achieved by Algorithm \ref{Algthm_AMSE_Opt} despite unguaranteed global optimality.
\vspace{-2mm}
\begin{algorithm}
\caption{Max. AMMSE Minimization}
\label{Algthm_AMSE_Opt}
\begin{algorithmic}[1]
\State \textbf{Initialize}: $n\gets 0$, $t_{0}^{(n)}\gets 0 $, $\bar{\alpha}_k^{(n)}\gets1 \ \forall k $
\Repeat
    \State $n\gets n+1$
    \State $t_{0}^{(n)} \gets \mathcal{A}_{lr}({P}_{0},\{\bar{\alpha}_{k}^{(n-1)}\}_{k=1}^{K})$
    \State $\{\mathbf{Q}_{k}^{(n)}\}_{k=1}^{K} \gets \arg \mathcal{A}_{lr}({P}_{0},\{\bar{\alpha}_{k}^{(n-1)}\}_{k=1}^{K})$
    \State $\mathbf{Q}^{(n)} \gets \sum_{i=1}^{i=K}\mathbf{Q}_{i}^{(n)}$
    \State \text{update} $\{\bar{\alpha}_{k}^{(n)}\}_{k=1}^{K}$ \text{using (\ref{Eq_Alpha_k})}
\Until{$|t_{0}^{(n)}-t_{0}^{(n-1)}|<\epsilon_{t}$ \text{or} $n=n_{\max}$ }
\end{algorithmic}
\end{algorithm}
\vspace{-2mm}
\section{Numerical Results}
\label{Section_Numerical Results}
We consider a BS equipped with $N_{t} = 4$ antennas serving $K=4$ single-antenna users. The $k$th user's CSIT consists of the first and second order statistics of $\mathbf{h}_{k}$, i.e. $\hat{\mathbf{h}}_{k} = \sigma_{c_{k}} [1,e^{j\varphi_{k}},\ldots ,e^{j(N_{t}-1)\varphi_{k}}]^{T} $ with $\varphi_{k} \sim \mathcal{U}(0,2\pi)$, and $\mathbf{R}_{e_{k}}=\sigma^{2}_{e_{k}}\mathbf{I}$. The per-antenna average path gain is defined as $\sigma^{2}_{k} = \sigma^{2}_{c_{k}}  + \sigma^{2}_{e_{k}}$.
This is similar to the model in \cite{Gonzalez-Coma2013} with an added feature of controlling individual CSIT qualities, e.g. $\sigma^{2}_{e_{k}}=0$ and $\sigma^{2}_{c_{k}}=0$ correspond to perfect and completely random CSITs for user $k$, respectively.
The average noise variance across users is given as $\sigma^{2}_{\text{av}}=\frac{1}{K}\Sigma_{k=1}^{K} \frac{\sigma^{2}_{n}}{ \sigma^{2}_{k}}$ from which the SNR is defined as $\frac{P_{t}}{K \sigma^{2}_{\text{av}}}$ \cite{Bogale2013}.
Throughout the simulations, all users are assumed to have unity path gains except for user-4 whose path gain is $3$ dB lower, i.e. $\sigma^{2}_{{1}} = \sigma^{2}_{{2}} = \sigma^{2}_{{3}} = 1$ and $\sigma^{2}_{{4}} = 0.5$. This is reflected in the CSIT quality (by influencing the UL pilot SNR or the feedback link capacity) yielding $\sigma^{2}_{e_{1}} = \sigma^{2}_{e_{2}} = \sigma^{2}_{e_{3}} = 0.05$ and $\sigma^{2}_{e_{4}}=0.1$. Therefore, user-4 is deemed the \emph{least fortunate}. $\sigma^{2}_{n}$ is fixed to $1$ where the SNR would only vary with $P_{t}$. For all simulation, $\{\hat{\mathbf{h}}_{k}\}_{k}^{K}$ are kept fixed where averaging is carried out over several independent realizations of $\{\tilde{\mathbf{h}}_{k} \}_{k=1}^{K}$. This corresponds to a scenario where the CSIT (and the optimized transmitter) remain unchanged over multiple (time or frequency) channel uses in which the channel response changes.
Similar performances are observed for different realizations of $\{\hat{\mathbf{h}}_{k}\}_{k}^{K}$.
\subsection{Power Minimization}
\begin{figure}[t!]\vspace{-5mm}
\centerline{
  \subfloat[AMMSE of user-4]{\label{Fig_P_AMSE}\includegraphics[width=1.88in]{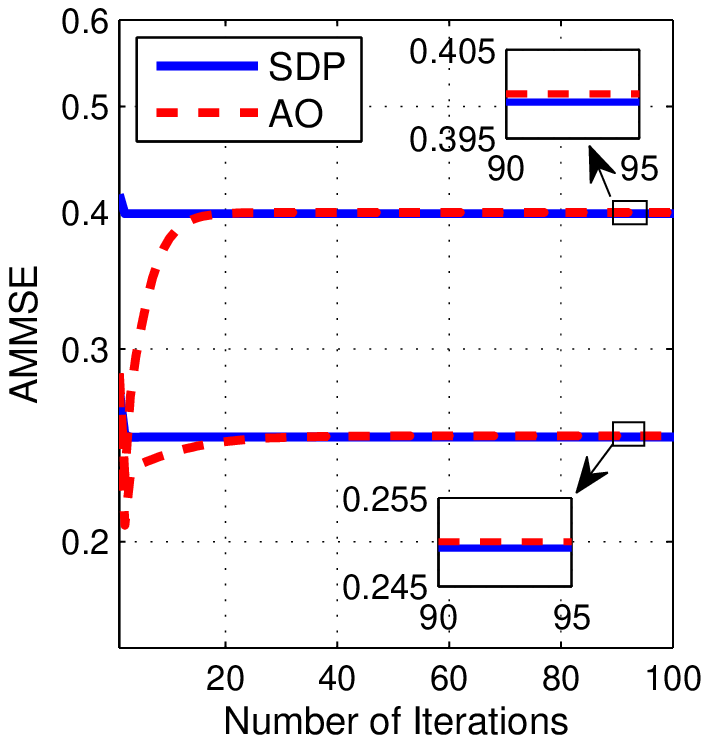}}
  \hspace{-3mm}
  \subfloat[Average SNR $(P_{t}/K\sigma^{2}_{\text{av}})$]{\label{Fig_Power} \includegraphics[width=1.88in]{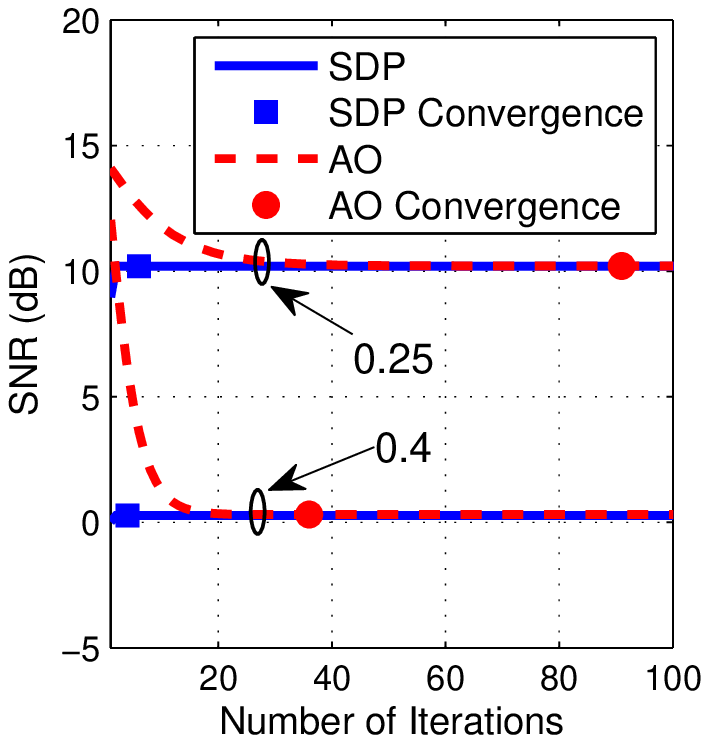}}
  }
  \label{Fig_Power_AMSE}
  \caption{$\mathcal{P}$: AO-Algorithm vs. SDP-Algorithm}
\end{figure}

The results obtained from solving $\mathcal{P}(0.25)$ and $\mathcal{P}(0.4)$ using Algorithm \ref{Algthm_Power_Opt} (SDP-Algorithm) are compared to those obtained using the AO-Algorithm \cite[Algorithm 1]{Gonzalez-Coma2013}.
The AO-Algorithm uses $4000$ Monte-Carlo realization and $\epsilon_{P}=10^{-4}$ \cite{Gonzalez-Coma2013} is used for both algorithms.
The AMMSE of the \emph{least fortunate} user and the total required SNR are plotted against the number of iterations in Fig. \ref{Fig_P_AMSE} and Fig. \ref{Fig_Power}, respectively.
Since the SDP-Algorithm does not involve Monte-Carlo integration, its AMMSE is obtained by averaging over the same realizations used for the AO-Algorithm.
Fig. \ref{Fig_P_AMSE} shows that both algorithms meet the AMMSE targets with high accuracy (the AMMSE approximation yields an error less than $0.5 \%$ for the SDP-Algorithm). Furthermore, it is evident from Fig. \ref{Fig_Power} that both algorithms yield similar minimized powers.
However, the SDP-Algorithm takes significantly less iterations to converge compared to the AO-Algorithm ($4$ vs. $36$ for $\mathcal{P}(0.4)$, and $6$ vs. $91$ for $\mathcal{P}(0.25)$). Contrary to the AO-Algorithm (which is randomly initialized), the first iteration of the SDP-Algorithm solves the problem using the first-order Taylor approximation of AMMSE yielding a good starting point for later iterations and consequently, reducing the total number of iterations.

For each iteration, the SDP-Algorithm has to solve a SDP with a
linear objective function, $K$ positive-semidefinite matrix variables of size $N_{t}\times N_{t}$ and $K$ linear inequality constraints.
Such problems can be efficiently solved using SDP solvers that apply Interior-Point methods (e.g. \cite{Sturm1999}) at a worst-case complexity cost that scales with $\mathcal{O}(K^{3.5}N_{t}^{6.5})$ \cite{Karipidis2008,Ye1997}. However, the actual runtime complexity scales far slower with $K$ and $N_{t}$. On the other hand, the complexity of the AO-Algorithm mainly comes from the numerical Monte-Carlo integration. In each iteration, multiple operations that scale with $\mathcal{O}( N_{t}^{2}KM)$, $\mathcal{O}( N_{t}K^{2}M)$ and $\mathcal{O}( N_{t}KM)$ are carried out, where $M$ is the number of Monte-Carlo realizations. For a practical system (e.g. $N_{t}\leq 8$), the per-iteration complexity of both algorithms is comparable and the actual runtime complexity for the SDP-Algorithm is significantly smaller.
\begin{figure}[t!]\vspace{-5mm}
\centerline{
  \subfloat[AMMSE of user-4]{\label{Fig_AMSE} \includegraphics[trim=0in 0in 0in 0.18in, clip=true, width=1.88in]{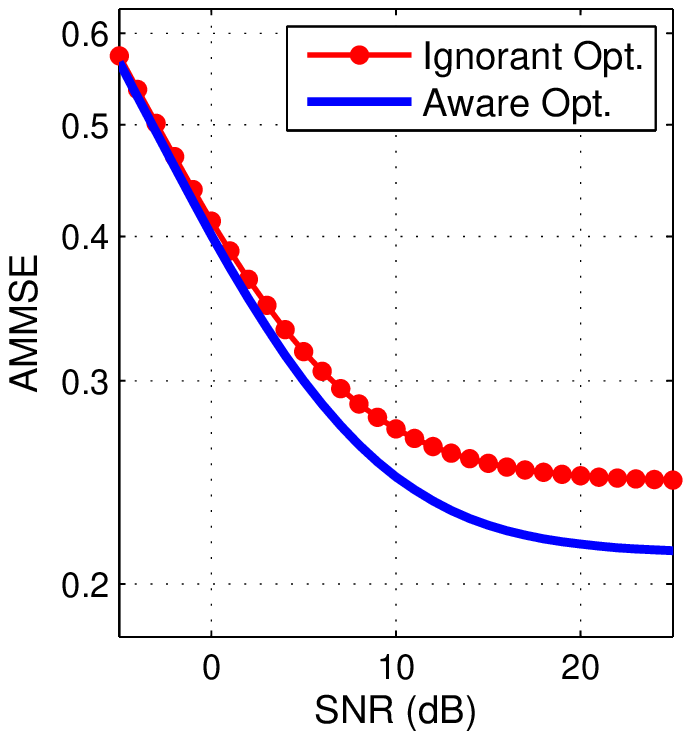}}%
    \hspace{-3mm}
  \subfloat[Average Rate of user-4]{\label{Fig_Rate} \includegraphics[trim=0in 0in 0in 0.18in, clip=true, width=1.88in]{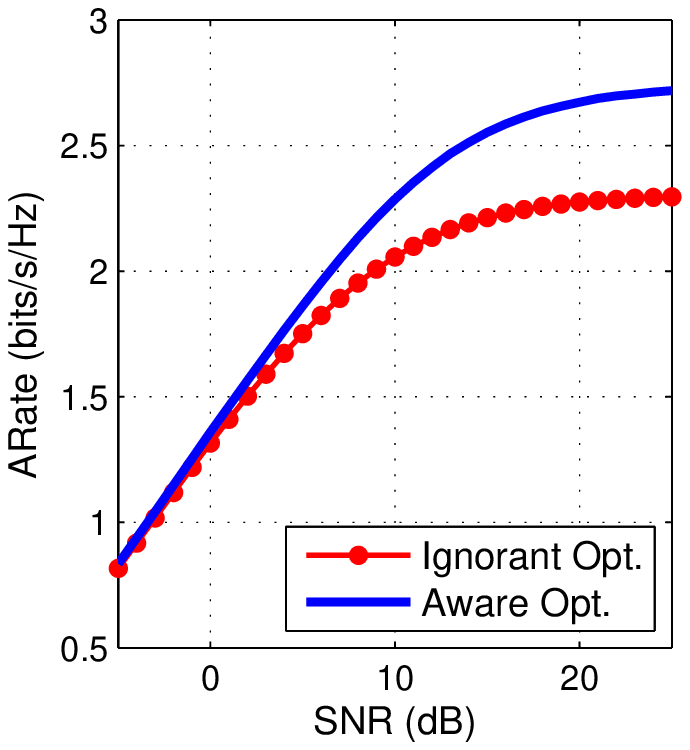}}
  }
  \label{Fig_AMSE_Rate}
    \caption{$\mathcal{A}$: Ignorant-Optimization vs. Aware-Optimization}
 \vspace{-5mm}
\end{figure}
\subsection{Maximum AMMSE Minimizations}
The robust CSIR-aware optimization proposed in Algorithm \ref{Algthm_AMSE_Opt} is compared to the ignorant optimization where the BS uses the available CSIT to jointly design the precoding vectors and the receivers which are forwarded to their corresponding users.
Results for the ignorant scheme are obtained using a slightly modified version of Algorithm \ref{Algthm_AMSE_Opt_Bisection}. Particularly, in step \ref{Algthm_AMSE_Opt_Bisection_Step_P}, the AMMSE constraints of $\mathcal{P}_{lr}$ are changed to $\hat{\varepsilon}_{k} \leq \bar{\varepsilon}_{0} \ \forall k$, and other changes are made accordingly.
The two schemes are compared in terms of the AMMSE and the Average Rate of the \emph{least fortunate} user in Fig. \ref{Fig_AMSE} and Fig. \ref{Fig_Rate}, respectively. The gain from the utilization of perfect CSIR grows with SNR as the channel estimation error becomes significant.
\section{Conclusion}
\label{Section_Conclusion}
In this paper, we proposed an AMMSE second order Taylor approximation for linearly-precoded MU-MISO systems with imperfect CSIT and perfect CSIR. This approximation was used to formulate the two fairness-based robust design problems: a power minimization problem ($\mathcal{P}$) and an AMMSE minimization problem ($\mathcal{A}$). Problem $\mathcal{P}$ was solved using a fast-converging algorithm based on recursive convex optimization.
On the other hand, problem $\mathcal{A}$ was solved by combining the recursive optimization approach with methods from \cite{Wiesel2006}. This CSIR-aware robust design was shown to have better performance compared to the ignorant design.
\appendix[]
\begin{proof}[Proof of Lemma \ref{Lemma_E_Complex_Quadratic_non_zero_mean}]
$\mathrm{E}\{Q_{1}Q_{2}\}$ could be written as the sum of 16 terms consisting of linear, quadratic, cubic and quartic forms of $\tilde{\mathbf{x}}$. By extending some of the identities in \cite[Ch. 8.2]{Petersen2008} (given for real Gaussian vectors) to ZMCSCG vectors, each term could be found individually and the expression in (\ref{Eq_E_Q1Q2}) is obtained.
\end{proof}

\bibliographystyle{IEEEtran}
\bibliography{References}

\begin{thebibliography}{10}
\providecommand{\url}[1]{#1}
\csname url@samestyle\endcsname
\providecommand{\newblock}{\relax}
\providecommand{\bibinfo}[2]{#2}
\providecommand{\BIBentrySTDinterwordspacing}{\spaceskip=0pt\relax}
\providecommand{\BIBentryALTinterwordstretchfactor}{4}
\providecommand{\BIBentryALTinterwordspacing}{\spaceskip=\fontdimen2\font plus
\BIBentryALTinterwordstretchfactor\fontdimen3\font minus
  \fontdimen4\font\relax}
\providecommand{\BIBforeignlanguage}[2]{{%
\expandafter\ifx\csname l@#1\endcsname\relax
\typeout{** WARNING: IEEEtran.bst: No hyphenation pattern has been}%
\typeout{** loaded for the language `#1'. Using the pattern for}%
\typeout{** the default language instead.}%
\else
\language=\csname l@#1\endcsname
\fi
#2}}
\providecommand{\BIBdecl}{\relax}
\BIBdecl

\bibitem{Viswanath2003}
P.~Viswanath and D.~Tse, ``{Sum capacity of the vector Gaussian broadcast
  channel and uplink-downlink duality},'' \emph{IEEE Transactions on
  Information Theory}, vol.~49, no.~8, pp. 1912--1921, Aug 2003.

\bibitem{Shenouda2008}
M.~Shenouda and T.~Davidson, ``{On the Design of Linear Transceivers for
  Multiuser Systems with Channel Uncertainty},'' \emph{IEEE Journal on Selected
  Areas in Communications}, vol.~26, no.~6, pp. 1015--1024, August 2008.

\bibitem{Bogale2013}
T.~Bogale and L.~Vandendorpe, ``{Linear Transceiver Design for Downlink
  Multiuser MIMO Systems: Downlink-Interference Duality Approach},'' \emph{IEEE
  Transactions on Signal Processing}, vol.~61, no.~19, pp. 4686--4700, Oct
  2013.

\bibitem{Joham2010}
M.~Joham, M.~Vonbun, and W.~Utschick, ``{MIMO BC/MAC MSE duality with imperfect
  transmitter and perfect receiver CSI},'' in \emph{IEEE 11th International
  Workshop on Signal Processing Advances in Wireless Communications (SPAWC)},
  June 2010, pp. 1--5.

\bibitem{Gonzalez-Coma2013}
J.~Gonzalez-Coma, M.~Joham, P.~Castro, and L.~Castedo, ``Power minimization in
  the multiuser downlink under user rate constraints and imperfect transmitter
  csi,'' in \emph{IEEE International Conference on Acoustics, Speech and Signal
  Processing (ICASSP)}, May 2013, pp. 4863--4867.

\bibitem{Vucic2009}
N.~Vucic and H.~Boche, ``{Robust QoS-Constrained Optimization of Downlink
  Multiuser MISO Systems},'' \emph{IEEE Transactions on Signal Processing},
  vol.~57, no.~2, pp. 714--725, Feb 2009.

\bibitem{Bashar2014}
M.~Bashar, Y.~Lejosne, D.~Slock, and Y.~Yuan-Wu, ``{MIMO broadcast channels
  with Gaussian CSIT and application to location based CSIT},'' in
  \emph{Information Theory and Applications Workshop (ITA)}, Feb 2014, pp.
  1--7.

\bibitem{Wiesel2006}
A.~Wiesel, Y.~Eldar, and S.~Shamai, ``{Linear precoding via conic optimization
  for fixed MIMO receivers},'' \emph{IEEE Transactions on Signal Processing},
  vol.~54, no.~1, pp. 161--176, Jan 2006.

\bibitem{Rice2009}
S.~H. Rice, ``The expected value of the ratio of correlated random variables,''
  Texas Tech University, 2009.

\bibitem{VanKempen2000}
G.~Van~Kempen and L.~Van~Vliet, ``{Mean and Variance of Ratio Estimators Used
  in Fluorescence Ratio Imaging},'' \emph{Cytometry}, vol.~39, no.~4, pp.
  300--305, 2000.

\bibitem{Boyd2004}
S.~P. Boyd and L.~Vandenberghe, \emph{{Convex Optimization}}.\hskip 1em plus
  0.5em minus 0.4em\relax Cambridge university press, 2004.

\bibitem{Luo2010}
Z.-Q. Luo, W.-K. Ma, A.-C. So, Y.~Ye, and S.~Zhang, ``{Semidefinite Relaxation
  of Quadratic Optimization Problems},'' \emph{IEEE Signal Processing
  Magazine}, vol.~27, no.~3, pp. 20--34, May 2010.

\bibitem{Gonzalez-Coma2013a}
J.~Gonzalez-Coma, M.~Joham, P.~Castro, and L.~Castedo, ``{Power minimization
  and QoS feasibility region in the multiuser MIMO broadcast channel with
  imperfect CSI},'' in \emph{IEEE 14th Workshop on Signal Processing Advances
  in Wireless Communications (SPAWC)}, June 2013, pp. 619--623.

\bibitem{Sturm1999}
J.~F. Sturm, ``{Using SeDuMi 1.02, a MATLAB toolbox for optimization over
  symmetric cones},'' \emph{{Optimization methods and software}}, vol.~11, no.
  1-4, pp. 625--653, 1999.

\bibitem{Karipidis2008}
E.~Karipidis, N.~Sidiropoulos, and Z.-Q. Luo, ``{Quality of Service and Max-Min
  Fair Transmit Beamforming to Multiple Cochannel Multicast Groups},''
  \emph{IEEE Transactions on Signal Processing}, vol.~56, no.~3, pp.
  1268--1279, March 2008.

\bibitem{Ye1997}
Y.~Ye, \emph{{Interior Point Algorithms: Theory and Analysis}}.\hskip 1em plus
  0.5em minus 0.4em\relax New York: John Wiley \& Sons, 1997.

\bibitem{Petersen2008}
K.~B. Petersen and M.~S. Pedersen, \emph{{The Matrix Cookbook}}, November 2008.

\end{thebibliography}

\end{document}